\documentclass[aps,prd,twocolumn]{revtex4-1}
\usepackage{graphicx,longtable,mathrsfs,color}
\usepackage[usenames,dvipsnames]{xcolor} 
\usepackage{amssymb,amsmath,mathtools,mathrsfs} 
\usepackage{epsfig,subfigure} 
\usepackage{booktabs,longtable} 
\usepackage{exscale,relsize}  
\usepackage[normalem]{ulem} 
\usepackage{enumerate}

\newcommand{\Hu}{{\cal H}}

\newcommand{\gamx}{{\Gamma_\chi}}
\newcommand{\1}{{^{(1)}}}

\newcommand{\kgal}{k_{\mathrm{gal}}}

\begin{document}

\title{New Gravitational Scales in Cosmological Surveys}

\author{Tessa Baker}
\email{tessa.baker@astro.ox.ac.uk}
\affiliation{Astrophysics, University of Oxford, DWB, Keble Road, Oxford OX1 3RH, UK}
\author{Pedro G. Ferreira}
\email{p.ferreira1@physics.ox.ac.uk}
\affiliation{Astrophysics, University of Oxford, DWB, Keble Road, Oxford OX1 3RH, UK}
\author{C. Danielle Leonard}
\email{danielle.leonard@astro.ox.ac.uk}
\affiliation{Astrophysics, University of Oxford, DWB, Keble Road, Oxford OX1 3RH, UK}
\author{Mariele Motta}
\email{mariele.motta@unige.ch}
\affiliation{D\'{e}partement de Physique Th\'{e}orique, Universit\'{e} de Gen\`{e}ve, 24 quai Ernest Ansermet,1211 Gen\`{e}ve 4, Switzerland}
\date{Received \today; published -- 00, 0000}

\begin{abstract}
\noindent In the quasistatic regime, generic modifications to gravity can give rise to novel scale-dependence of the gravitational field equations. Crucially, the detectability of the new scale-dependent terms hinges upon the existence of an effective mass scale or length scale at which corrections to GR become relevant. Starting from only a few basic principles, we derive the general form of this scale-dependence. Our method recovers results previously known in the specific case of Horndeski gravity, but also shows that they are valid more generally, beyond the regime of scalar field theories. We forecast the constraints that upcoming experiments will place on the existence of a new fundamental mass scale or length scale in cosmology.
\end{abstract}

%\pacs{98.80.-k, 98.80.Es, 95.36.+d, 95.36.+x}

\maketitle

\section{Introduction}
\label{section:intro}
Our current working hypothesis is that the dominant force acting on large scales in the universe is gravity, and that gravity is accurately described by General Relativity (GR). In practice, we often focus on cosmological systems that are smaller than the Hubble scale; this permits us to make a set of simplifications referred to as the quasi-static (QS) limit. 

 The QS regime corresponds to a window of lengthscales that are considerably smaller than the cosmological horizon, such that ${\cal H}/k\ll1$, but sufficiently large such that linear perturbation theory is still valid.
In concrete figures, a reasonable estimate would be distances less than $600h^{-1}$ Mpc but greater than $10-20h^{-1}$ Mpc. The usual argument is that within this window of lengthscales the time derivatives of metric potentials are significantly smaller than their spatial derivatives. In GR this statement is a natural consequence of the subhorizon condition ($\Hu/k\ll1$), since the linear gravitational potentials evolve on the Hubble timescale. In practical terms, implementing conditions such as \mbox{$|\ddot\Phi|\ll |\nabla^2\Phi|$} makes the linearised field equations easy to work with.

When we go over to a modified theory of gravity, the situation is less clear-cut. On one hand we can reason that any gravity theory consistent with current observations must behave in a manner very similar to $\Lambda$CDM, so we expect our quasistatic limit to be preserved. On the other hand, when we modify GR we naturally introduce new dynamical degrees of freedom (hereafter d.o.f.) which might have evolutionary timescales different from $\Hu$. In this paper we will assume that the new d.o.f. are sufficiently subdominant that a QS limit still exists for most of the history of the universe.

The largest distances that can be probed by current and next-generation galaxy surveys fall predominantly within the QS regime. To use these experiments to test the laws of gravity, we need to understand the typical behaviour of non-GR  theories in the QS limit. There is a long-standing intuition that modified gravity theories generically lead to a novel dependence of observables on the length scale at which they are measured, e.g. the density-weighted growth rate, $f\sigma_8(z)$, becomes a function of wavenumber $k$. Work has already begun to search for such signatures \cite{Johnson2014}.

The goal of this paper is to make concrete these intuitions. What are the implications of a (non-)detection of scale-dependence for the host of  gravity theories in the current literature \cite{Clifton2012}? What are the most theoretically-motivated parameters for observers to measure? We present three main results:

\vspace{2mm}
\noindent 1) Considering a frequently-used parameterisation of the linearised gravitational field equations, we show that only a few key physical principles are needed to derive the fixed scale-dependence of many gravity theories in the QS regime. Our results apply to any theory with second-order equations of motion and one new spin-0 degree of freedom, which does \textit{not} have to be a scalar field. This generalises the results of \cite{defelice2011, Unobservables2013, Gleyzes2013, Bloomfield2013} beyond single-scalar field theories (see \cite{Silvestri2013} for related ideas). The derivation is compact and does not require knowledge of a gravitational action -- hence its generality.
\vspace{2mm}

\noindent 2) We show that the detectability of this non-GR scale-dependence hinges crucially on the existence of new physical quantities (characteristic masses, lengths, etc.) that are generically introduced when modifying GR. If all such parameters are tuned to be comparable to the Hubble scale, it is highly unlikely that any novel scale-dependence of observables will be detectable in the QS regime.
\vspace{2mm}

\noindent 3) Turning these ideas around, we isolate the leading-order scale-dependent terms and estimate the constraints placed upon them by next-generation cosmological experiments. The headline results are displayed in Fig.~\ref{fig:ellipses}.

The structure of this paper is as follows: in \S\ref{section:set-up} we discuss how characteristic physical quantities enter most popular theories of gravity. In \S\ref{section:derivation} we derive the result described in 1) above. In \S\ref{section:mass_scales} we discuss the implications of a (non-)detection of scale dependence, i.e. point 2) above. We also isolate the leading-order contribution to scale-dependence; in \S\ref{section:forecasts} we carry out example forecasts for future constraints on this leading-order term (point 3) above). We conclude in \S\ref{section:conclusions}. Some technical details are relegated to the appendices.

\section{New Scales \& Set-up}
\label{section:set-up}
New physical scales are a near-universal feature of alternative gravity theories. This is no surprise: the success of GR in describing the Solar System generally forces us to introduce a `transition scale' into gravitational physics, positing that gravity reduces to GR on one side of this transition scale, but receives modifications on the other side. 

Let us elaborate with some examples. Probably the most familiar example of a new scale arises in scalar-tensor theories, where a mass scale emerges from second derivatives of the potential, $V(\phi)_{,\phi\phi}$. In $f(R)$ gravity the new scale is more often thought of as a Compton wavelength for the scalaron, but is similarly derived from derivatives of an effective potential (a function of $f_{,RR}$) \cite{Pogosian2008}. New scales can arise in a different way when there is non-trivial coupling between the matter energy-momentum tensor and the scalar degree of freedom, e.g. in theories which display chameleon screening, the transition scale is marked by a potential well depth, $|\Phi|\sim 10^{-6}$ \cite{Davis2012,Jain2013}.

Vector-tensor theories are often endowed with an energy (mass) scale at which violations of Lorentz invariance become manifest. This new scale can be an explicit parameter in the Lagrangian of the theory (as, for example, in Horava-Lifschitz gravity \cite{Horava2009, Sotiriou_HL_review}), or it can arise implicitly via spontaneous Lorentz violation at the level of the field equations (e.g. in the effective field theory of a vector coupled to gravity \cite{Gripaios2007}).

It is well known that current bimetric theories not only have an explicit mass scale (the mass of the graviton), but also a system-dependent length scale, the Vainshtein radius, which signals the onset of screening \cite{Babichev2013}. 
Similarly, higher-dimensional theories such as DGP \cite{Deffayet2002} and other braneworld models can have both an explicit scale such as a warp factor or crossover scale, as well as Vainshtein radii.

Finally, there has been recent interest in non-local theories \cite{Maggiore22014, Dirian2014, Maggiore2014, Foffa2014} containing Lagrangian terms such as $R\,\square^{-2}R$. The solutions for $\square^{-1}R$ involve integrals over Green's functions for the $\square^{-1}$ operator, $G(x,x^\prime)$. This naturally suggests a characteristic scale between the spacetime points $x$ and $x^\prime$ over which non-local interactions occur.

In essence, new transition scales can be dependent on a variety of physical quantities such as energy, ambient density, acceleration, potential, etc. Throughout this paper we will be agnostic about the origin of any new transition scale. In many gravity theories the new scale(s) are tuned to be of order the Hubble scale today, in the hope that they might replace the cosmological constant as the driver for accelerated expansion. In other theories, however, an effective cosmological constant is included in the theory (sometimes explicitly, sometimes via a `back door') \cite{Faulkner2007,HuSawicki}. Then, since cosmic acceleration is already taken care of, new mass or length scales inherent in the theory may take a much wider range of values.

Let us define a wavenumber $\kgal$ that represents the largest perturbation mode that can be reasonably well-measured by next-generation cosmological surveys. Let us also introduce a mass scale $M$ that represents the transition scale accompanying some generic modifications to gravity; in what follows we will also frequently interpret $M$ as an inverse lengthscale. (Note that we can extract a mass or length scale from any of the physical quantities discussed above by using appropriate factors of $c,\,\hbar$ etc., and taking appropriate powers).

We can then envisage three scenarios:
\begin{enumerate}[a)]
\item $M\sim \Hu\Rightarrow M\ll \kgal$ in the QS regime. 
\item $\Hu \ll M\lesssim \kgal$. 
\item $\kgal\ll M$.  
\end{enumerate} 
In this paper we will treat situations a) and b). We will work with one new dynamical degree of freedom, denoting its perturbations by $\chi$. $\chi$ does \textit{not} have to be a scalar field, but could instead be a spin-0 perturbation of a new vector or tensor field, a new d.o.f. excited in the metric (i.e. a metric d.o.f. which is non-dynamical in GR), a Stuckelberg field, or several other possibilities \cite{PPF2013}. 

For the purposes of this paper we are only interested in the relative orders of magnitude of terms, not in precise factors of order unity. We can therefore write time derivatives of $\chi$ as $\dot\chi=\gamx\chi$, where $\gamx$ is the evolutionary timescale of the new d.o.f. perturbation. There are two possibilities for this timescale: it could either be approximately the Hubble timescale (like for the metric potentials), or it could be a new, shorter timescale determined by $M$ (note that with appropriate factors of $c$ a mass is dimensionally equivalent to an inverse timescale). For now we will maintain generality, but later on we will we see that setting $\gamx\sim \Hu$ or $\gamx\sim M$ can have different consequences.

We will not consider situation c), because in this scenario the existence of a quasistatic limit becomes questionable. If $M$ is very large and $\gamx\sim M$, then we have that $\dot\chi$ is very rapidly evolving, violating quasistaticity. Furthermore, the novel effects that occur at wavelengths close to the lengthscale $M^{-1}$ are likely to occur inside the nonlinear regime, which is beyond the scope of this paper. 

Consideration of the Friedmann equation suggests that the background (zeroth-order) values of any new fields present are constrained to evolve on Hubble timescales. Using the example of a scalar field to illustrate, we mean that $\dot{\phi}^2\sim \Hu^2\phi^2$. Now one might well argue that if a field evolves like $\Hu$ on the background, its perturbations must evolve like $\Hu$ too, that is, only the case $\gamx\sim\Hu$ is of interest. In \S\ref{section:mass_scales} we will argue that if this is true, it seems unlikely that the scale-dependent properties of modified gravity will be measurable any time soon. 

\section{Derivation}
\label{section:derivation}
To obtain result 1) of \S\ref{section:intro}, we first need to take a step back to the gravitational field equations. A linear combination of two components of the tensor field equations gives the gravitational Poisson equation, whilst the transverse spatial component gives the `slip' relation (shown here at late times; our conventions for the metric potentials are given in Appendix~\ref{app:derivs}):
\begin{align}
-k^2\Phi&=4\pi G\mu(a,k) a^2{\bar \rho}_m \Delta_m \label{eqNP}\\
\Phi&=\gamma(a,k) \Psi \label{gamma}
\end{align}
where $\bar{\rho}_m$ is the mean matter density, $\Delta_m$ is the gauge-invariant density contrast, and a sum over all matter species is implied.

In the expressions above we have introduced two functions, $\mu(a,k)$ and $\gamma(a,k)$, that have been used extensively as a parameterisation of modified field equations in the QS regime \cite{BZ2008, Bean2010, Hojjati2013, Simpson2013, Motta2013}. This parameterisation is convenient for theoretical work because it parameterises the `raw' gravitational field equations obtained directly from the action. However, a slightly different parameterisation (`$\left\{\tilde{\mu},\,\Sigma\right\}$' -- see eq.(\ref{convert_params})) is preferred for data analyses, because it leads to minimal parameter degeneracy when combining redshift-space distortions and weak lensing surveys. For this reason we will present our theoretical results in terms of $\left\{\mu,\,\gamma\right\}$, then rotate to the basis  $\left\{\tilde\mu,\,\Sigma\right\}$ for the forecasts in \S\ref{section:forecasts}.

To manipulate a particular gravity theory into the form of eqs.(\ref{eqNP}) and (\ref{gamma}), one begins with the full (ie. unparameterised) Poisson equation and slip relation of the theory, and the linearly perturbed equation of motion for the new d.o.f. The steps are as follows:
\begin{enumerate}
\item Apply the quasistatic approximation to the metric terms in the three equations listed above, that is, drop terms containing $\ddot\Phi,\,\dot\Phi,\,\ddot\Psi$ ad $\dot\Psi$. Of the remaining terms, discard those with prefactors that evolve on Hubble timescales, i.e. drop $\Hu^2\Phi$, but keep $\nabla^2\Phi$. Time derivatives of $\chi$ should be replaced by $\dot\chi\approx\gamx\chi$ and $\ddot\chi\approx\gamx^2\chi$ but not discarded.
\item Take two linear combinations of the unparameterised slip equation and the equation of motion for the d.o.f.: one combination that eliminates $\chi$, and one that eliminates $\Psi$. The form of $\gamma(a,k)$ can be read off from the first linear combination. 
\item Substitute the ratios $\Psi/\Phi$ and $\chi/\Phi$ obtained in the previous step into the right-hand side of the Poisson equation, so that it is written purely in terms of $\Phi$ (plus the usual GR term in $\Delta$). Rearrange this equation into the form of eq.(\ref{eqNP}) and read off $\mu(a,k)$.
\end{enumerate}

We wish to avoid laboriously carrying out these steps for many individual gravity theories. So instead we will apply this procedure to a set of `template' field equations that reflect the structure of real theories. A similar derivation was presented first in \cite{Silvestri2013}; the addition we make is the explicit consideration of a new mass scale, $M$, as discussed in \S\ref{section:set-up}.

We will write down these templates in the conformal Newtonian gauge. To maintain transparent correspondence with the usual linearised Einstein equations we will not use an explicitly gauge-invariant combination of variables to represent the new d.o.f.. However, we will use the fact that the equations must ultimately have a gauge-invariant formulation to guide the construction of our templates.

 For example, the usual gauge-invariant Bardeen potentials $\hat\Phi$ and $\hat\Psi$ contain first- and second-order time derivatives respectively. This means that $\hat\Psi$ can only appear in the Poisson equation as part of the combination $\dot{\hat\Phi}+\Hu\hat\Psi$ (in which the second time-derivatives cancel out -- see Appendix~\ref{app:derivs} or \cite{PPF2013}) to avoid converting the Poisson equation from a constraint into a dynamical equation.

An example will help to clarify this point and allow us to introduce some notation. For the case where the dimensionless new d.o.f. $\chi$ is a scalar field or a fluid energy density, the Poisson equation has the form:
\begin{align}
-2k^2\Phi&=8\pi G a^2\bar{\rho}_m\Delta_m+\Phi\left(h_1k^2+h_2\left[\Hu^2,M^2,\Hu M\right]\right)\nonumber\\
&+h_3\left[\Hu,M\right]\dot\Phi+m_2\left[\Hu^2,M^2,\Hu M\right]\Psi\label{Poissonscalar}\\
&+\chi\left(g_1k^2+g_2\left[\Hu^2,M^2,\Hu M\right]\right)+g_3\left[\Hu,M\right]\dot\chi\nonumber
\end{align}
where $M$ is the potential new mass scale. Throughout this paper we will use notation like $\left[\Hu^2, M^2, \Hu M \right]$ to indicate a function of time which has dimensions of mass-squared. Terms appearing in this function can have three possible order-of-magitudes: $\Hu^2$, $M^2$ or $\Hu M$. The numerical coefficients accompanying these order-of-magnitude terms are unimportant for our purposes. We will use the dimensionless order-unity coefficients $h_i$, $g_i$ etc. simply as a convenient way to refer to individual terms. In complete analogy, $\left[\Hu,M\right]$ denotes a time-dependent function with the dimension of mass, which can have two possible orders of magnitude: $\sim\Hu$ or $\sim M$.

Note that $\Psi$ appears up to one derivative order lower than $\Phi$ in eq.(\ref{Poissonscalar}). This is due to the aforementioned requirement that it must be possible to `repackage' these terms into the combination \mbox{$\alpha\dot\Phi+\beta(\dot\Phi+\Hu\Psi)$}, where $\alpha$ and $\beta$ are numerical coefficients.
 
Carrying out step 1 of our procedure, eq.(\ref{Poissonscalar}) becomes:
 \begin{align}
-2k^2\Phi&=8\pi G a^2\rho\Delta+\Phi\left[h_1k^2+h_2M^2\right]+\Psi\left[m_2M^2\right]\nonumber\\
&+\chi\left[g_1k^2+g_2M^2+g_3M\gamx\right]\label{QSPoissonscalar}
\end{align}

For brevity we will not write here the non-quasistatic templates for the slip relation and equation of motion, for this scalar field/fluid example; they are given in eqs.(\ref{eomscalar}) and (\ref{slipscalar}). We move straight to their QS limits, which are:
\begin{align}
\chi\Big[d_1 \gamx^2&+d_2 M\gamx+d_3 M^2+d_4 k^2\Big]\label{QSeomscalar}\\
&+\Phi\left[b_2 M^2+b_3k^2\right] +\Psi\left[c_2 M^2+c_3k^2\right]=0\nonumber\\
&\Phi-\Psi=e_0\Phi+j_0\Psi+f_0\chi& \label{QSslipscalar}
\end{align}

\noindent Carrying out steps 2 and 3 described above, we obtain the following forms for $\mu$ and $\gamma$:
\begin{align}
\gamma&=\frac{p_1+p_2\frac{M^2}{k^2}+p_3\frac{\gamx M}{k^2}+p_4\frac{\gamx^2}{k^2}}{q_1+q_2\frac{M^2}{k^2}+q_3\frac{\gamx M}{k^2}+q_4\frac{\gamx^2}{k^2}}\label{gamscalar}\\
\mu&=\left[p_1+p_2\frac{M^2}{k^2}+p_3\frac{\gamx M}{k^2}+p_4\frac{\gamx^2}{k^2}\right]\times\nonumber\\
\Big[t_1&+t_2\frac{M^2}{k^2}+t_3\frac{\gamx M}{k^2}+t_4\frac{\gamx^2}{k^2}+t_5\frac{M^4}{k^4}+t_6\frac{\gamx M^3}{k^4}\nonumber\\
&+t_7\frac{\gamx^2M^2}{k^4}\Big]^{-1}\label{muscalar}
\end{align}
where the $p_i,\,q_i$ and $t_i$ are simple algebraic combinations of the order-unity coefficients in eqs.(\ref{QSPoissonscalar})-(\ref{QSslipscalar}). Their precise forms are given in Table~\ref{tab:scalar} in Appendix~\ref{app:derivs}.

We immediately recognise that eqs.(\ref{gamscalar}) and (\ref{muscalar}) subsume some results already known for scalar field theories \cite{defelice2011, Unobservables2013, Silvestri2013, Gleyzes2013, Bloomfield2013}, but note that we have not needed to use the complex form of the Horndeski Lagrangian \cite{Horndeski, Deffayet2011,Gleyzes2014} to obtain them here. For example, we see that $\mu$ and $\gamma$ share the same numerator; an equivalent result was proved in \cite{Silvestri2013, Gleyzes2013} (note that other authors use a slightly different parameterisation variables to ours, equivalent to the set $\left\{\tilde{\mu},\,\gamma\right\}$, where $\tilde\mu$ is defined in eq.(\ref{convert_params})).
  
Our expression for $\mu$ contains three terms -- $t_5$, $t_6$ and $t_7$ -- which have not been included in works focused on Horndeski theories. In all Horndeski-type theories we have seen investigated this $k^{-4}$ dependence of $\mu$ and $\gamma$ is not present, because the terms represented by $h_2$, $m_2$, $g_2$ and $g_3$ in eq.(\ref{QSPoissonscalar}) do not feature in the QS limit of their field equations \footnote{Without the presence of a new mass scale, these terms are of magnitude $\sim \Hu^2$ and hence are discarded when the QS limit is taken.}. However, we will leave these terms in our general expressions because they could exist in some as-yet-undiscovered gravity theory. It is in the spirit of this paper to remain as agnostic as possible. 

We stress that these forms for $\mu$ and $\gamma$ have been derived using purely a few basic principles, such as gauge-invariance and restriction to second-order equations of motion. We have used neither a model-specific action nor a general EFT-inspired one. In fact, if we repeat steps 1-3 for the case where the new d.o.f. is the spatial spin-0 perturbation of a timelike vector field (i.e. $|\chi|\sim |k|V$), such as occurs in Einstein-Aether theories \cite{Jacobson2008,Zlosnik2008}, we reach exactly the same form as eqs.(\ref{gamscalar}) and (\ref{muscalar}). The derivation is given in Appendix~\ref{app:derivs}, and only differs from the one shown here in small details. 

Note also that the authors of \cite{Solomon2014,Koennig} have recently derived expressions equivalent to $\mu$ and $\gamma$ in a particular bigravity model, and found them to have a form contained by eqs.(\ref{gamscalar}) and (\ref{muscalar}) -- see eqs.(85) and (86) of \cite{Koennig}. This is not unexpected, since bigravity -- despite its complex field equations (not fully encapsulated by eqs.~\ref{QSPoissonscalar}-\ref{QSslipscalar}) -- ultimately has the same physical features as the cases treated explicitly here, i.e. it introduces only one new spin-0 degree of freedom and respects gauge-invariance and locality. If the current stability issues surrounding generic bigravity models \cite{Comelli2014} can be resolved, then eqs.(\ref{gamscalar}) and (\ref{muscalar}) can be used as a universal parameterisation for virtually \textit{all} theories with second-order equations of motion and a single d.o.f of any type.

 \section{REGIMES OF INTEREST} 
 \label{section:mass_scales}
As discussed in \S\ref{section:set-up}, there are essentially two choices for the scale $M$ that we have introduced. Either we can set $M$ to be of order the Hubble scale, or we can posit a new scale, $M\gg\Hu$, which marks the transition from the GR limit to some larger theory of gravity. This choice governs whether it makes sense to invest effort searching for scale-dependent signatures of modified gravity \cite{Johnson2014}.
 \vspace{5mm}
 
\noindent \textit{Case a).} Many gravity theories explicitly tune their new scale(s) to be of order \mbox{$H_0\sim10^{-3}$} eV in order to produce a viable expansion history. For example, in recent bigravity models the graviton mass is taken to be of order the present Hubble scale. In this case there is no choice other than setting $M=\gamx=\Hu$. If we carry out the full QS approximation \textit{all} scale-dependent terms in eqs.(\ref{gamscalar}) and (\ref{muscalar}) must be discarded and we are left with only the simple time-dependent expressions. Scale-dependence in $\mu$ and $\gamma$ only arises if we consider the first-order corrections in $(\Hu/k)^2$ to the QS approximation, which leads to \footnote{In case a) the precise form of the $p_i$, $q_i$ and $t_i$ differ in very small details from those given in Table~\ref{tab:scalar}, e.g. factors of two. Given the approximate nature of this work it is not necessary to display another complete set of tables with the modified coefficients.}:
 \begin{align}
 \gamma(a)&=\frac{p_1+\left(p_2+p_3+p_4\right)\frac{\Hu^2}{k^2}}{q_1+\left(q_2+q_3+q_4\right)\frac{\Hu^2}{k^2}} \\
  \mu(a)&=\frac{p_1+\left(p_2+p_3+p_4\right)\frac{\Hu^2}{k^2}}{t_1+\left(t_2+t_3+t_4\right)\frac{\Hu^2}{k^2}}
 \end{align}
 Expressions of precisely this form have been worked out explicitly for numerous theories \cite{defelice2011, Solomon2014, Koennig}.
 
In this scenario any scale-dependence of observables will be very weak. Until we are able to survey a substantial fraction of our Hubble volume, it is arguably better to focus our efforts on tightly constraining the time dependence of $\{\mu,\gamma\}$ or $\{\tilde\mu,\Sigma\}$ by combining information from all scale bins. 
 \vspace{5mm}

 \noindent \textit{Case b).} In this scenario one posits a new physical transition scale $M^{-1}$ below the cosmological horizon distance, such that $M\gg\Hu$. For example, in $f(R)$ gravity the new mass scale is approximately given by \mbox{$M^2\propto 1/f_{,RR}$}, and $f_{,RR} R\ll1\,\Rightarrow\,1/f_{,RR} \gg \Hu^2$ is needed to ensure stability in the matter-dominated era \cite{Hojjati2012}. 
   
Recall that in \S\ref{section:set-up} we introduced a wavenumber $\kgal$ that typified the maximum distance scale that could be well-constrained by near-future galaxy surveys.  Given the lack of deviations from $\Lambda$CDM+GR to date, one might naively assume the maximum value for $M$ is of order $\kgal$. In principle one should then attempt to constrain the full form of eqs.(\ref{gamscalar}) and (\ref{muscalar}). However, it has been shown that in practice it will be difficult to constrain all the individual $p_i,\,q_i$ and $t_i$ \cite{Hojjati2014}.

This motivates us to consider a slightly less accurate but simpler approach. We perform a Taylor expansion of eqs.(\ref{gamscalar}) and (\ref{muscalar}) in the vicinity of the naive assumption $M\lesssim \kgal$, and keep only the leading order terms. We show in Appendix~\ref{appendix:As_and_Ms} that this gives the following expressions:
\begin{align}
{\mu}(a,k)&\simeq1+A_{\mu}(a)\left[1+\left(\frac{M_{\mu}(a)}{k}\right)^2\right] \label{iii}\\ 
\gamma(a,k)&\simeq1+A_\gamma(a)\left[1+\left(\frac{M_\gamma(a)}{k}\right)^2\right]\label{kkk}
\end{align}
The precise content of $M_\mu(a)$ and $M_{\gamma}(a)$ depends on whether $\gamx\sim M$ or $\gamx \sim \Hu$, but this kind of detail is not important here. The thrust of our argument is that equations (\ref{iii}) and (\ref{kkk}) provide a simple, general and theoretically well-motivated description of scale-dependence that should be easily applicable to observations. They include case a) as a limit, if $M_\mu(a)$ and $M_{\gamma}(a)$ are set to be of order $\Hu$.
 
 \begin{figure}[t!]
\begin{center}
\hspace{-2mm}
\includegraphics[scale=0.27]{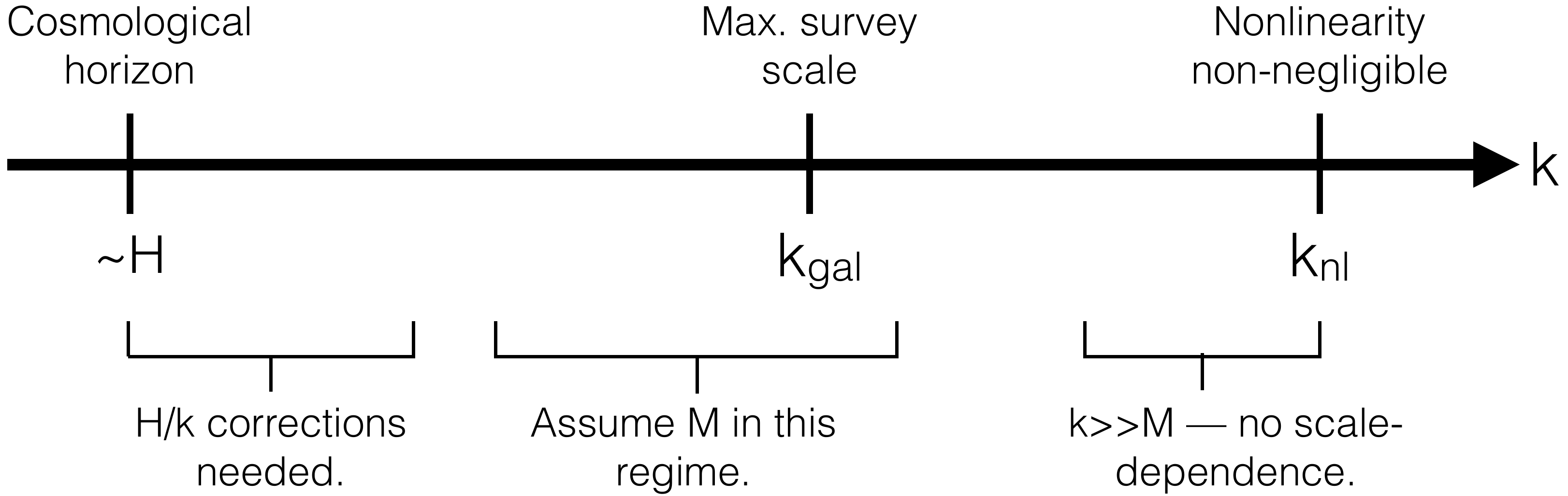}
\caption{Schematic diagram illustrating the arguments of \S\ref{section:mass_scales} (case b).}
\label{figure:scales_line_plot}
\end{center}
\end{figure}
\vspace{5mm}
 
 From the discussion of this paper, we now understand that a non-zero detection of $M_\gamma$, $M_\mu$, $A_\mu$ or $A_\gamma$ would signify one of two possible things:
 \begin{enumerate}
 \item A breakdown of the QS approximation $\Hu/k \ll 1$. The scale-dependence would then be due to first-order corrections in $\Hu^2/k^2$.
 \item The existence of a new scale in gravitational physics \footnote{At the risk of over-emphasis, remember from our discussion above that the converse is not true. A lack of observed scale-dependence does not rule out modified gravity.}.
 \end{enumerate}
 Either of these scenarios would have profound implications for our understanding of gravity on large scales.

\section{Detecting scale dependence.}
\label{section:forecasts}
We now speculate on the constraints that can be placed on the kind of parameterisation introduced above with future cosmological surveys. As we mentioned in \S\ref{section:derivation}, the function set $\left\{\mu,\,\gamma\right\}$ is the most convenient for theoretical work. However, a change of basis will enable us to minimise parameter degeneracies when using redshift-space distortion (hereafter RSD) and weak lensing data \cite{Simpson2013,Linder2013}. We introduce the new function set $\left\{\tilde\mu,\,\Sigma\right\}$, related to the old set by:
\begin{align}
\tilde\mu(a,k)&=\frac{\mu(a,k)}{\gamma(a,k)} & \Sigma(a,k)&=\frac{\mu(a,k)}{2}\left(1+\frac{1}{\gamma(a,k)}\right)
\label{convert_params}
\end{align}
Effectively, $\tilde\mu$ parameterises the geodesic equation for non-relativistic particles that governs the linear collapse of cold dark matter; $\Sigma$ parameterises the geodesic equation for photons that governs weak gravitational lensing. However, it is important to note that gravitational lensing is also sensitive to $\tilde\mu$, because the lensing convergence and shear spectra involve integrals over the matter power spectrum, which is affected by modified structure growth \cite{Simpson2013,Leonardinprep}.

We will write the new function set in a form analogous to eqs.(\ref{iii}) and (\ref{kkk}), that is:
\begin{align}
{\tilde{\mu}}(a,k)&\simeq1+A_{\tilde{\mu}}(a)\left[1+\left(\frac{M_{\tilde{\mu}}(a)}{k}\right)^2\right] \label{mutilde} \\ 
\Sigma(a,k)&\simeq1+A_\Sigma(a)\left[1+\left(\frac{M_\Sigma(a)}{k}\right)^2\right] \label{Sigma}
\end{align}
We stress that we are more interested in the general form of the scale-dependence rather than the precise (and lengthy) expressions relating $A_{\tilde\mu},\,A_\Sigma, M^2_{\tilde\mu}$, and $M^2_\Sigma$ to the coefficients of the field equations  (though for completeness the relationships between the $\left\{\mu,\,\gamma\right\}$ and $\left\{\tilde{\mu},\,\Sigma\right\}$ parameterisations are given in Appendix \ref{appendix:conv}).

An unavoidable feature of model-independent tests of gravity is that ansatzes must be chosen for the time-dependent functions. There must be enough parameters in the ansatz to capture important signatures in the data without weakening the constraints too severely. As a simplicity-motivated test case, we will choose our ansatz to be (partially following \cite{Simpson2013}):
\begin{align}
A_{\tilde\mu}(a)&=\tilde{\mu}_0\frac{\Omega^{GR}_\Lambda(a)}{\Omega^{GR}_{\Lambda 0}} \label{Amutilde} \\
A_\Sigma(a)&=\Sigma_0\frac{\Omega^{GR}_\Lambda(a)}{\Omega^{GR}_{\Lambda 0}} \label{ASigma}\\
M_{\tilde\mu}&=m_{\tilde\mu}\,(20 H_0) \label{yrt}\\
M_{\Sigma}&=m_{\Sigma}\,(20 H_0)\label{yrt2}
\end{align}
where $m_{\tilde\mu}$ and $m_{\Sigma}$ are constants. Remembering that we will want to be able to interpret the $M_i^{-1}$ as lengthscales, it is convenient to introduce a subhorizon distance unit of $(20H_0)^{-1}$ and express $M_i^{-1}$ in units of this distance. In the simple forecasts here we will focus on perturbative observables, fixing the background expansion history to match that of the $\Lambda$CDM+GR model and using Planck best-fit cosmological parameters \cite{Planck_params}. For an analysis that accounts for a modified expansion history see \cite{Leonardinprep}.

In principle we should really allow $M_{\tilde\mu}$ and $M_\Sigma$ to be functions of time. Treating them as constants simply corresponds to imposing the same overall time-dependent amplitudes $A_i(a)$ on both the scale-free and scale-dependent modifications to the QS field equations.

The set of four parameters that we will forecast for is:
\begin{align}
\tilde{\mu}_0,\;\;\Sigma_0,\;\;\tilde{\mu}_0m^2_{\tilde\mu},\;\;\Sigma_0m^2_\Sigma
\end{align}
Note that the scale-dependent parts of the parameterisation are sensitive to a degenerate combination of the time-dependent amplitude and the possible new effective mass/length scale; we cannot constrain $M_{\tilde\mu}$ and $M_\Sigma$ individually. 
\begin{center}
\begin{figure*}[t]
\hspace{-1.2cm}
\subfigure{\label{fig:mu0Sig0}\includegraphics[scale=0.475]{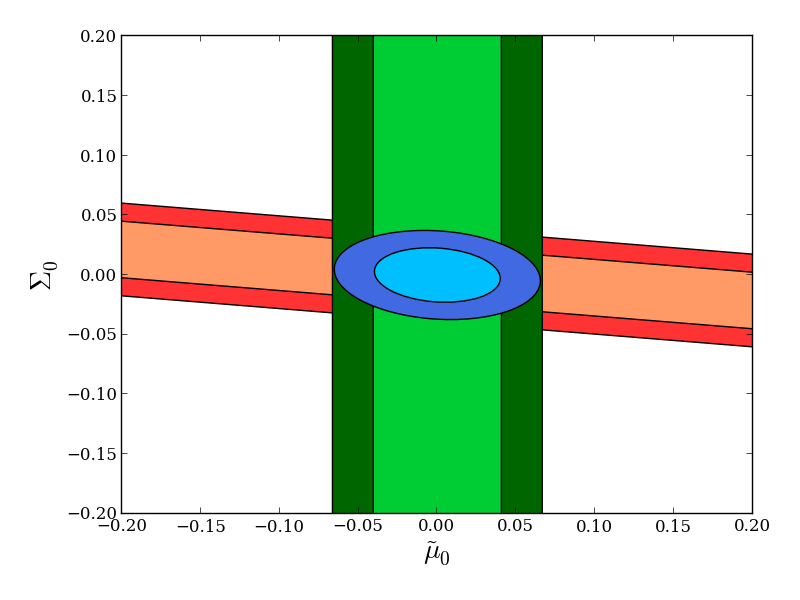}}
\hspace{-0.5cm}
\subfigure{\label{fig:MmuMSig}\includegraphics[scale=0.476]{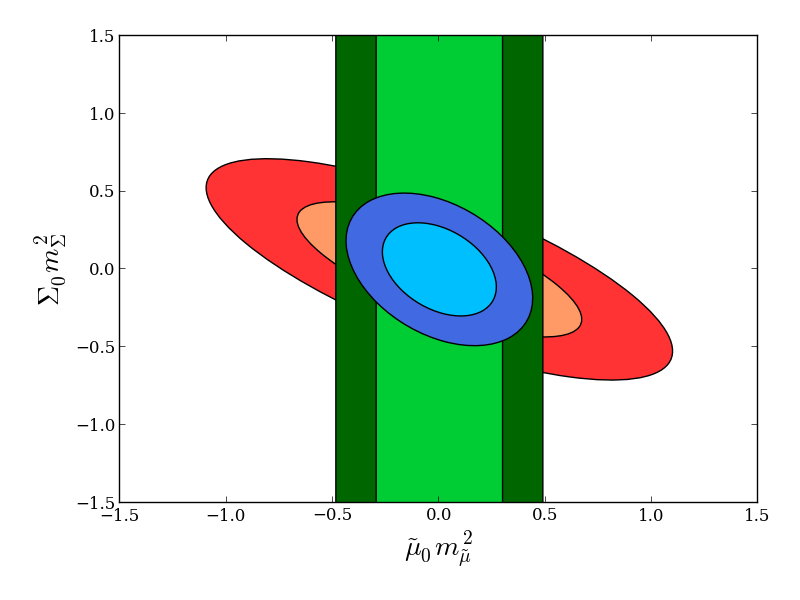}}%
\vspace{-0.8cm}%
\caption{Forecast constraints on the scale-independent (left panel) and scale-dependent (right panel) parts of the $\{\tilde{\mu},\Sigma\}$ parameterisation, using a DETF stage 4-like experiment. Redshift-space distortions (green contours) constrain only the two parameters in the $\tilde\mu(z,k)$ function, $\tilde{\mu}_0$ and $m_{\tilde\mu}$. Gravitational weak lensing (red contours) predominantly constrain the parameters in $\Sigma(z,k)$, but also have some dependence on $\tilde{\mu}_0$. Blue contours show the combined constraints. The parameters $m_{\tilde{\mu}}$ and $m_{\Sigma}$ are $M_{\tilde{\mu}}$ and $M_\Sigma$ expressed in distance units of $\left(20 H_0\right)^{-1}$\,Mpc, see eqs.(\ref{yrt}) and (\ref{yrt2}).}
\label{fig:ellipses}
\end{figure*}
\end{center}

We consider a Dark Energy Task Force stage 4 (DETF4)  experiment that combines a galaxy clustering survey and a dedicated tomographic weak lensing survey. Weak lensing utilises scale-dependent information naturally, as the standard quantities to calculate are angular power spectra. RSD measurements, however, generally do not. Usually we talk about the density-weighted growth rate, $f\sigma_8(z)$, implicitly assuming data from all scales ($k$-bins) has been combined.

We modify this situation by dividing each redshift bin of our hypothetical survey into five bins in $k$-space, with edges \mbox{$\left[0.005,0.02,0.05,0.08,0.12,0.15\right]$h\,Mpc$^{-1}$}; the choice of $k$-binning is analogous to \cite{Johnson2014}, and the upper limit is chosen to cut off before nonlinearities start to dominate. It seems likely that as our survey sizes increase large-scale measurements will improve, whilst small-scale measurements will remain dominated by a lack of understanding of baryonic physics and the effects of nonlinearities. For this reason we will take our $k$-bins to have the following fractional errors at all redshifts, from large-scale to small-scale:~$\left[0.01,0.03,0.03,0.09,0.09\right]$. 

For the tomographic gravitational lensing, we consider five source bins. These are constructed by taking the total distribution of source galaxies as:
\begin{equation}
n(z) \propto z^{\alpha} e^{-\left(\frac{z}{z_0}\right)^\beta}
\label{nofz}
\end{equation}
with $\alpha=2$, $\beta=1.5$, and $z_0=z_m/1.412$ where $z_m$ is the median redshift of the survey \cite{Thomas2009, Amendola2008}.  $n(z)$ is then divided into five bins between $z=0.5$ and $z=2$, each with equal numbers of galaxies.  The lensing errors are encoded in the covariance matrices:
\begin{align}
{\bf C}_{ij}(\ell)=\sqrt{\frac{2}{(2\ell+1)f_{sky}}}\left( P_{ij}^{\kappa,\,GR}(\ell)+\delta_{ij}\,\frac{\langle\gamma_{\mathrm{int}}^2\rangle}{\bar{n}_i}\right)
\end{align}
 where $P_{ij}^{\kappa,\,GR}(\ell)$ is the cross-correlated convergence power spectrum sourced by galaxies in  bins $i$ and $j$, and $f_{sky}$ is the fraction of the sky covered by the survey. $\langle\gamma_{\mathrm{int}}^2\rangle^{\frac{1}{2}}$ is the r.m.s. intrinsic shear and $\bar{n}_i$ is the number of galaxies per steradian in source bin $i$. For further details see \cite{Hu1999, Leonardinprep}.

Our model-agnostic approach to the field equations (eqs.\ref{QSPoissonscalar}-\ref{QSslipscalar}) means that factors of order unity are of no relevance here. For this reason we do not attempt a detailed, experiment-specific forecast (for which the $k$-bin errors and maximum $k$ value would evolve with redshift). More precise forecasts can be found in, for example, \cite{Amendola_Fogli2013, Taddei2014}. Other model-independent tests of $\Lambda$CDM using the growth rate have recently appeared in \cite{Nesseris_Sapone}.

Fig.~\ref{fig:ellipses} shows marginalised 2D constraints on the scale-independent and scale-dependent parts of the parameterisation (eqs.~\ref{mutilde} and \ref{Sigma}). RSDs (green contours) constrain only $\tilde{\mu}_0$ and $\tilde{\mu}_0M^2_{\tilde\mu}$, whilst weak lensing (red contours) is sensitive to all four parameters. The authors of \cite{Simpson2013} have applied a scale-independent parameterisation to CFTHLens+WiggleZ data, finding that lensing is only weakly sensitive to ${\tilde\mu}_0$. We agree with these scale\textit{-independent} results, but find that the scale\textit{-dependent} parts of the `lensing function' $\Sigma(a,k)$ and the `RSD function' $\tilde{\mu}(a,k)$ are more strongly correlated \cite{Leonardinprep}, see the right panel of Fig.~\ref{fig:ellipses}.

 A rough estimate of the precision with which we will be able to measure these new effective mass/length scales in cosmology gives us
$\sigma \left(m^2_{\tilde\mu}\right)\sim{\sigma\left(\tilde{\mu}_0 m^2_{\tilde\mu}\right)}/{\sigma\left(\tilde{\mu}_0 \right)}\sim 6.7$ and 
$\sigma \left(m^2_{\Sigma}\right)\sim{\sigma\left(\Sigma_0 m^2_{\Sigma}\right)}/{\sigma\left(\Sigma_0 \right)}\sim 13.2$. 
Interpreting these limits as distance scales, we find lower bounds of order $ M^{-1}_{\tilde\mu}\geq 364\,\mathrm{Mpc}$ and $M^{-1}_{\Sigma}\geq  260\, \mathrm{Mpc}$. 
We see that RSDs are  more sensitive than weak lensing to new fundamental scales. That is, we should be able to pin down a new characteristic distance scale all the way up to $364$~Mpc with growth rate measurements.

Given that the bounds we have found on the $M_i$ are comparable to $k_{\mathrm{gal}}$, the Taylor expansion of eqs.(\ref{gamscalar}) and (\ref{muscalar}) (see Appendix~\ref{appendix:As_and_Ms}) may not be accurate enough. Yet, the well-behaved form of eqs.(\ref{gamscalar}) and (\ref{muscalar}) (ratios of quadratic polynomials) suggest that subsequent corrections in higher powers of $M^2/k^2$ might change the bounds placed on $M_{\tilde\mu}$ and $M_\Sigma$ by, at most, a factor of a few.

\section{Conclusions}
\label{section:conclusions}

The spirit of this work has been to take a step back from detailed model-specific investigations in alternative theories of gravity. Ultimately, the expressions collected in Appendix~\ref{app:derivs} link the parameterisation we presented in eqs.(\ref{mutilde}) and (\ref{Sigma}) to the field equations of a specific gravity theory; but, as we hope has been clear, this is not the strategy we are advocating.  Instead, the goal of this paper has been to highlight the fact that even the exotic plethora of gravity theories on the market today share basic physical features which endow them with the same structure in the quasistatic regime. 

We have found that the scale-dependence of gravity theories is closely linked to an effective mass scale or length scale which the linearised field equations inherit from their parent quadratic action. Our derivation has allowed for the evolutionary timescale of the new degree of freedom to be affected by this new scale in the system, rather than relying too heavily on GR-based intuitions that might suggest $\dot\chi$ is negligible in the QS regime.

In many theories a new mass scale is tuned to be $\sim\Hu$ in order to produce accelerated expansion. Indeed, the motivation behind much of the current bestiary of modified gravity theories is to render the cosmological constant obsolete. {\it Generally} this renders the scale-dependendence undetectable in the quasistatic regime. We conjecture that most of the theories giving rise to detectable scale-dependence are those which introduce a new scale much larger than the Hubble scale ($M\gg\Hu$), and meanwhile rely on a cosmological constant to achieve a viable expansion history.

We caution, though, that theories with screening mechanisms complicate the issue somewhat. One can envisage a smaller, plausibly detectable length scale emerging if it is a compound of fundamental scales and couplings to the energy-momentum tensor.

Note that at no point in this paper have we needed a concrete action from which to start our calculations: knowledge of the basic physical properties of a theory (eg. second-order equations of motion and a single dynamical spin-0 perturbation) is sufficient. We have trivially recovered results of \cite{defelice2011,Gleyzes2013, Bloomfield2013, Unobservables2013}, and have found them to be more general than previously realised (see \cite{Silvestri2013} for a similar analysis along these lines).

We advocate that measurement of scale-dependent observables is an important and feasible target for next-generation cosmology experiments. They have the potential to unveil a scale at which new physics beyond $\Lambda$CDM+GR kicks in. More conservatively, scale-dependent measurements would also act as an essential test of the quasistatic approximation that has rapidly grown in popularity over the past few years.

\textit{Acknowledgments.} 
It is a pleasure to thank L. Amendola, J. Gleyzes, M. Kunz, M. Lagos, L. Miller, J. Noller, F. Piazza, A. Silvestri, F. Vernizzi and H. Winther. TB is supported by All Souls College, Oxford. PGF acknowledges support from Leverhulme, STFC, BIPAC and the Oxford Martin School. DL is supported by the Rhodes Trust. MM acknowledges support from the Swiss National Science Foundation and the Balzan foundation via the University of Oxford.

\begin{widetext}
\appendix

\section{Derivation of $\mu$ and $\gamma$.}
\label{app:derivs}
In this appendix we show explicitly how the forms of eqs.(\ref{gamscalar}) and (\ref{muscalar}) were reached.

\subsection{Simple Scalar/Fluid Case}

Consider a theory of a single scalar field with second-order equations of motion. For this to be a valid theory of gravity, we know that the linearised equation of motion (hereafter e.o.m.) of the scalar must have a gauge-invariant formulation. It must therefore be possible to group all terms in the e.o.m. into gauge-invariant combinations; there cannot be any gauge-varying terms `left over' after this regrouping has happened.

More explicitly: let us write the perturbed line element in a general gauge as (recall $\epsilon=\nu=0$ in the conformal Newtonian gauge):
\begin{align}
\label{line_element}
ds^2 =& a(\eta)^2\Big[-(1+2 \Psi)d\eta^2-2(\vec{\nabla}_i\epsilon)d\eta\,dx+\left(1-2\Phi\right)\gamma_{ij}+\left(D_{ij}\nu\right) \,dx^i dx^j \Big]
\end{align}
where $D_{ij}=\vec{\nabla}_i\vec{\nabla}_j-\frac{1}{3}\delta_{ij}\vec{\nabla}_k\vec{\nabla}^k$. A gauge-invariant combination containing the scalar field perturbation is (hats signify gauge-invariant variables):
\begin{align}
\label{giscalar}
\hat{\delta\phi}=\delta\phi+\frac{\dot\phi}{\Hu}(\Phi+k^2\nu)
\end{align}
This means that, because the linearised e.o.m. contains a term in $\delta\ddot\phi$, it must also contain $\ddot\Phi$ so that the two can be packed together (along with other terms) as $\ddot{\hat{\delta\phi}}$.

An additional subtlety surrounds the Newtonian potential $\Psi$. The gauge-invariant version of $\Psi$ is one of the well-known Bardeen variables:
\begin{align}
\hat\Psi&=\Psi-\frac{1}{2}(\ddot\nu+2\dot\epsilon)-\frac{1}{2}\Hu (\dot\nu+2\epsilon) \label{Psi_def}
\end{align}
Note that the combination above contains a second-order time derivative, $\ddot\nu$. This is potentially dangerous: if $\dot{\hat\Psi}$ or $\ddot{\hat\Psi}$ appeared in the equations of motion, they would generically introduce the Ostrogradski instability \cite{Ostrogradski}.

Yet one sees $\dot\Psi$ appearing all the time in field equations of gravity theories. How can this be? A careful examination of field equations reveals that it always appears accompanied by terms in $\ddot\Phi$ and $\dot\Phi$, such that they can be regrouped into the following combination and its time derivative:
\begin{align}
\label{alphadef}
\hat\alpha&=\dot{\hat\Phi}+\Hu\hat\Psi=\dot\Phi+\Hu\Psi+\frac{1}{2}(\dot\Hu-\Hu^2)(\dot\nu+2\epsilon)
\end{align}
where $\hat\Phi$ is the other standard Bardeen variable:
\begin{align}
\hat\Phi&=\Phi-\frac{1}{6}k^2\nu+\frac{1}{2}\Hu(\dot\nu+2\epsilon) 
\end{align}
The dangerous second time derivative has been eradicated from eq.(\ref{alphadef}). Therefore $\dot{\hat\alpha}$ can appear in a second-order e.o.m. without causing any instabilities. Similarly, the combination $\hat\alpha$ can appear in a constraint (first-order) equation.

Now, when we view equations in the conformal Newtonian gauge we do not `see' the $\nu$ or $\epsilon$ terms, but the above arguments still control the structure of the e.o.m.s. We have deduced that $\Psi$ must always present at one derivative order \textit{lower} than $\Phi$, and a brief glance at the perturbed Horndeski equations in \cite{defelice2011} confirms that this is indeed always the case in scalar field theories. 

Using these ideas we can write down a general template for the e.o.m. We make use of dimensional consistency, and the fact that the only objects with dimensions of mass we have to work with are $\Hu$ (recall we are setting $c=1$) and the new scale in our theory, $M$. The result is (where the notation implied by square brackets is explained below eq.(\ref{Poissonscalar}) in the main text):
\begin{align}
d_1 \ddot{\delta\phi}&+d_2 [\Hu,M]\,\dot{\delta\phi}+d_3 [\Hu^2,M^2,\Hu M] {\delta\phi}+d_4 k^2\,{\delta\phi}+b_0\ddot{\Phi}+b_1[\Hu,M]\dot{\Phi}+b_2[\Hu^2,M^2,\Hu M]\Phi+b_3 k^2 \,\Phi\nonumber\\
&+c_1[\Hu,M]\dot\Psi+c_2[\Hu^2,M^2,\Hu M]\Psi+c_3 k^2\Psi=0\label{eomscalar}
\end{align}
In fact this template holds not only for a scalar field, but also for the fractional energy density of a fluid or effective fluid, whose gauge-invariant version is:
\begin{align}
\hat\delta=\delta-(1+w)\left(3\Phi-\frac{1}{2}k^2\nu\right)
\end{align}
This means that eq.(\ref{eomscalar}) is also valid for a generic dark fluid or DGP gravity (in which the new d.o.f. in the 4D effective theory can be treated as perturbations of a radiation-like `Weyl fluid'). 

Moving on to the linearised gravitational field equations themselves, similar logic applies. However, the Poisson equation is a constraint equation and therefore can only contain $\dot{\delta\phi}$, $\dot\Phi$ and $\Psi$, plus undifferentiated ${\delta\phi}$ and $\Phi$. This leads us to the template of eq.(\ref{Poissonscalar}), which we reproduce here for convenience:
\begin{align}
-2k^2\Phi&=8\pi G a^2\rho\Delta+\Phi\left(h_1k^2+h_2[\Hu^2,M^2,\Hu M]\right)+h_3[\Hu,M]\dot\Phi+m_2[\Hu^2,M^2,\Hu M]\Psi\nonumber\\
&+{\delta\phi}\left(g_1k^2+g_2[\Hu^2,M^2,\Hu M]\right)+g_3[\Hu,M]\dot{\delta\phi}
\label{fna}
\end{align}
The transverse spatial Einstein equation has no time-derivative terms, because it already has dimensions of mass squared from the spatial derivatives:
\begin{align}
k_ik_j\left(\Phi-\Psi\right)&=k_i k_j\left(e_0\Phi+j_0\Psi+f_0{\delta\phi}\right),\quad\quad\mathrm{where}\quad i\neq j\label{slipscalar}
\end{align}
We usually pull off the spatial derivatives to obtain the slip relation of eq.(\ref{QSslipscalar}).

When we apply the quasistatic approximation, $M$ dominates over $\Hu$ in all the coefficient brackets; the result is eqs.(\ref{QSPoissonscalar})-(\ref{QSslipscalar}). The straightforward algebraic steps outlined in \S\ref{section:derivation} then lead to the expressions for $\mu$ and $\gamma$ in eqs.(\ref{gamscalar}) and (\ref{muscalar}). The coefficients in eqs.(\ref{gamscalar}) and (\ref{muscalar}) are combinations of those in eqs.(\ref{eomscalar}), (\ref{fna}) and (\ref{slipscalar}) as shown below:
\begin{center}
\begin{tabular}{|c|c|}\hline
{\bf Coeff.} & {\bf Relation to field equations} \\ \hline 
$p_1$ & $f_0c_3-(1+j_0)d_4$ \\ \hline
$p_2$ & $f_0c_2-(1+j_0)d_3$ \\ \hline
$p_3$ & $-(1+j_0)d_2$ \\ \hline
$p_4$ & $-(1+j_0)d_1$ \\ \hline
\end{tabular}
\hspace{1cm}
\begin{tabular}{|c|c|}\hline
{\bf Coeff.} & {\bf Relation to field equations} \\ \hline 
$q_1$ & $(e_0-1)d_4-{b}_3f_0$ \\ \hline
$q_2$ & $(e_0-1)d_3-{b}_2f_0$ \\ \hline
$q_3$ & $(e_0-1)d_2$ \\ \hline
$q_4$ & $(e_0-1)d_1$ \\ \hline
\end{tabular}
\end{center}
\begin{center}
\begin{tabular}{|c|c|}\hline
{\bf Coeff.} & {\bf Relation to field equations} \\ \hline 
$s_1$ & $(e_0-1)c_3-{b}_3(1+j_0)$ \\ \hline
$s_2$ & $(e_0-1)c_2-{b}_2(1+j_0)$ \\ \hline
$t_1$ & $p_1\left(1+\frac{{h}_1}{2}\right)-\frac{1}{2}g_1s_1$ \\ \hline
$t_2$ & $p_2\left(1+\frac{{h}_1}{2}\right)+\frac{1}{2}({h}_2p_1+m_2q_1-g_1s_2-g_2s_1)$ \\ \hline
$t_3$ & $p_3\left(1+\frac{{h}_1}{2}\right)-\frac{1}{2}g_3s_1$ \\ \hline
$t_4$ & $p_4\left(1+\frac{{h}_1}{2}\right)$ \\ \hline
$t_5$ & $\frac{1}{2}({h}_2p_2+m_2q_2-g_2s_2)$ \\ \hline
$t_6$ & $\frac{1}{2}({h}_2p_3+m_2q_3-g_3s_2)$ \\ \hline
$t_7$ & $\frac{1}{2}({h}_2p_4+m_2q_4)$ \\ \hline
\end{tabular}
\label{tab:scalar}
\end{center}

\subsection{Vector-like Case}
What about gravity theories where the new d.o.f. comes not from a scalar field, but a vector field? For example, in Einstein-Aether gravity there is a single new spin-0 perturbation $V$ contained within the spatial part of a timelike vector field: $\delta A_i=1/a (\nabla_i V)$. The appropriate gauge-invariant version of $V$ is reminiscent of the scalar field case (a general algorithm for finding such gauge-invariant field combinations was given in \cite{PPF2013}):
\begin{align}
\label{giV}
\hat{V}&=V-\frac{1}{\Hu}\left(\Phi-\frac{k^2\nu}{6}\right)
\end{align}
However, the difference here is that $V$ has dimensions of inverse mass. This affects the terms and coefficient dimensions that can appear in the e.o.m., Poisson equation and slip relation. The full (non-QS) versions are are shown below:
\begin{align}
d_1 \ddot V&+d_2 [\Hu,M]\,\dot V+d_3 [\Hu^2,\Hu M] V+d_4 k^2\,V+b_0\frac{\ddot\Phi}{\Hu}+
b_1\dot\Phi+b_2[\Hu,M]\Phi+b_3\frac{k^2}{\Hu}\Phi\nonumber\\
&+c_2\dot\Psi+c_3[\Hu,M]\Psi=0\label{eomvec}\\
-2k^2\Phi&=\kappa a^2\rho\Delta+\Phi\left(h_1k^2+h_2[\Hu^2,M^2,\Hu M]\right)+\dot\Phi\left(h_3[\Hu,M]+h_4\frac{k^2}{\Hu}\right) \nonumber\\
 &+\Psi\left(m_2[\Hu^2,M^2,\Hu M]+m_4k^2\right)
 +g_1[\Hu]k^2V+g_3k^2\dot V\label{Poissonvec}\\
\Phi-\Psi&=e_0\Phi+e_1\frac{\dot\Phi}{\Hu}+j_0\Psi+f_0\left[\Hu\right] V+f_1\dot V\label{slipvec}
\end{align}
There are several terms in the equations above that were not present in the scalar field/fluid case, namely $b_0$, $b_3$, $e_1$ and $h_4$. This is due to the denominator in eq.(\ref{giV}) -- we need to add these terms to make sure that the combination $\hat V$ and its derivatives can be formed \footnote{The analogous expressions for $\hat{\delta\phi}$ involve $\dot\phi/\Hu\sim \phi$ (since background variables evolve at the Hubble rate), so the denominator has no affect there.}.

There are also some possible terms missing, e.g. we have not allowed a term proportional to $M^2 V$ to appear in eq.(\ref{eomvec}). If this was present, the requirement of a gauge-invariant formulation means that we would also need to have a term proportional to $M^2\Phi/\Hu$ for it to partner with. If $M\sim\Hu$, this has already been accounted for in eq.(\ref{eomvec}). If $M\gg\Hu$ then such terms dominate the equations and force $\mu\rightarrow 0$, $\gamma\rightarrow 0$. 
 
 Similar considerations, after carrying out steps 1-3 described in \S\ref{section:derivation}, indicate that for the vector case we require $\Gamma_V\sim\Hu$ to avoid the situation $\mu\rightarrow 0$, $\gamma\rightarrow 0$. This greatly reduces the number of terms that survive to the final expressions, which are (using overbars just to avoid confusion with Table~\ref{tab:scalar}):
 \begin{align}
 \gamma&=\frac{\bar p_1+\bar p_2\frac{M^2}{k^2}}{\bar q_1+\bar q_1\frac{M^2}{k^2}}\\
 \mu&=\frac{\bar p_1+\bar p_2\frac{M^2}{k^2}}{\bar t_1+\bar t_2\frac{M^2}{k^2}+\bar t_5\frac{M^4}{k^4}}
\end{align} 
 where
 \begin{center}
\begin{tabular}{|c|c|}\hline
{\bf Coeff.} & {\bf Relation to field equations} \\ \hline 
$\bar p_1$ & $-(1+j_0)d_4$ \\ \hline
$\bar p_2$ & $c_3f_0$ \\ \hline
$\bar q_1$ & $(e_0+e_1-1)d_4$ \\ \hline
$\bar q_2$ & $-b_2f_0$ \\ \hline
$\bar t_1$ & $\bar p_1\left(1+\frac{h_1+h_4}{2}\right)+\frac{1}{2}m_4\bar q_1+\frac{1}{2}g_1b_3(1+j_0)$ \\ \hline
$\bar t_2$ & $\bar p_2\left(1+\frac{h_1+h_4}{2}\right)+\frac{1}{2}(h_2\bar p_1+m_2\bar q_1+m_4\bar q_2)$ \\ \hline
$\bar t_5$ & $\frac{1}{2}({h}_2\bar p_2+m_2\bar q_2)$ \\ \hline
\end{tabular}
\label{tab:vector}
\end{center}

 We conclude that, apart from some pathological cases, eqs.(\ref{gamscalar}) and (\ref{muscalar}) act as a universal form for theories with one spin-0 degree of freedom and second-order e.o.m.s.
 
 \section{Expansion of $\{\mu,\,\gamma\}$. }
 \label{appendix:As_and_Ms}
 
 In this appendix we show how the parameterisation of eqs.(\ref{iii}) and (\ref{kkk}) is obtained from the expansion of eqs.(\ref{gamscalar}) and (\ref{muscalar}). For ease of notation we define $y=\left(M/k\right)^2$, and perform a Taylor expansion about the point $y\rightarrow 0$. Effectively, $y\rightarrow 0$ corresponds to very small scales inside both the cosmological horizon and the new lengthscale $M^{-1}$, where $\mu$ and $\gamma$ are virtually scale-independent. We are expanding $\mu$ and $\gamma$ `upwards' in distance scales from the full QS limit, to find the first scale-dependent corrections that occur. (Of course, when taken far enough, the limit $y\rightarrow 0$ enters the nonlinear regime. Implicitly we stop before this point, i.e. we are taking $y\rightarrow \epsilon$, a very small number).
  
 To simplify the expressions, we will present here the case where $\gamx\sim \Hu$. The case with $\gamx\sim M$ is analogous but more algebraically cumbersome.
  
 For $y<1$ the Taylor expansion of $\gamma(a,k)$ yields (suppressing arguments of conformal time):
 \begin{align}
 \gamma(y)&\approx \gamma(y=0)+\gamma^\prime |_{y=0} y+{\cal O}\left(y^2\right) \label{ooo}\\
\gamma(y)&\approx\frac{p_1}{q_1}+\frac{p_1}{q_1}\left[\frac{p_2}{p_1}-\frac{q_2}{q_1}\right]y\\
&=\frac{p_1}{q_1}\left(1+\left[\frac{p_2}{p_1}-\frac{q_2}{q_1}\right]\frac{M^2}{k^2}\right)
\end{align}
It is convenient to separate out the GR limit explicitly by writing $p_1/q_1=1+A_\gamma$:
\begin{align}
\gamma(y)&=1+A_\gamma\left\{1+\left[\frac{p_2}{p_1}-\frac{q_2}{q_1}\right]\left(\frac{1+A_\gamma}{A_\gamma}\right)y\right\}
\end{align}
which is our desired form with:
\begin{align}
M^2_\gamma=\left[\frac{p_2}{p_1}-\frac{q_2}{q_1}\right]\left(\frac{1+A_\gamma}{A_\gamma}\right)\,M^2=\left[\frac{p_2}{p_1}-\frac{q_2}{q_1}\right]\frac{p_1}{p_1-q_1}\,M^2 \label{yyy}
\end{align}
The expressions for $A_\mu$ and $M_\mu$ are entirely analogous, with the simple replacement $q_i\rightarrow t_i$.
 
 One may be concerned that, given the fairly small lower bound found on the lengthscale $M^{-1}$ in \S\ref{section:forecasts}, we cannot guarantee that the entire extent of a galaxy survey satisfies the condition $M \leq k$. However, eq.(\ref{yyy}) makes it clear that the true lengthscale ($M^{-1}$) and the parameter we constrain ($M_\gamma^{-1}$) are related by an unknown factor. A factor of order unity here would be enough to push the true scale above the reach of galaxy surveys, so that the condition $M\leq k$ is always satisfied. The error associated with dropping the higher-order terms in eq.(\ref{ooo}) will not change our estimates by orders of magnitude, which is the only precision we are aiming for in the generalised analysis of this paper.
 
 \section{Conversion of $\{\mu,\,\gamma\}$  to $\{\tilde\mu,\,\Sigma\}$. }
 \label{appendix:conv}

 We reproduce here the relationship between the $\{\mu,\,\gamma\}$ theory-convenient parameterisation and the  $\{\tilde\mu,\,\Sigma\}$ observations-convenient one. 
 \begin{align}
\tilde\mu(a,k)&=\frac{\mu(a,k)}{\gamma(a,k)} & \Sigma(a,k)&=\frac{\mu(a,k)}{2}\left(1+\frac{1}{\gamma(a,k)}\right)
\end{align}
We can write $\{\tilde\mu,\,\Sigma\}$ in the form of eqs.(\ref{mutilde}) and (\ref{Sigma}). The relationship to the coefficients used in the $\{\mu,\,\gamma\}$ basis (eqs.~\ref{iii}-\ref{kkk}) is not particularly illuminating, but we give it here for completeness (suppressing the time argument throughout):
 \begin{align}
 A_{\tilde\mu}&=\frac{A_\mu-A_\gamma}{1+A_\gamma} & A_\Sigma&=\frac{1}{2}\left(A_\mu+A_{\tilde\mu}\right)\label{Atildemu}
 \end{align}
 \begin{align}
 M^2_{\tilde\mu}&=\frac{A_\mu M_\mu^2\left(1+A_\gamma\right)-A_\gamma M_\gamma^2\left(1+A_\mu\right)}{\left(A_\mu-A_\gamma\right)\left(1+A_\gamma\right)}\label{Mmutilde}
 \\
 M^2_\Sigma&= \frac{A_\mu M_\mu^2+A_{\tilde\mu} M_{\tilde\mu}^2}{A_\mu+A_{\tilde\mu}} \label{MSigma}
 \end{align}

 \appendix
\end{widetext}

\bibliographystyle{apsrev4-1}
\vskip -0.28in
%\bibliography{refs_DL}


\end{document}